\documentstyle[aps,prb,epsf,multicol]{revtex}
\begin{document}
\title{Charge correlations in the weakly doped $t-J$ model
       calculated by projection technique}
\author{Matthias Vojta and Klaus W. Becker}
\address{Institut f\"{u}r Theoretische Physik,
Technische Universit\"{a}t Dresden, D-01062 Dresden, \\ Germany}
\maketitle

\begin{abstract}
We study frequency- and wave-vector dependent charge correlations in weakly
doped antiferromagnets using Mori-Zwanzig projection technique.
The system is described by the two-dimensional $t$-$J$ model.
The ground state is expressed within a cumulant
formalism which has been successfully applied to study magnetic properties
of the weakly doped system. Within this approach the ground state
contains independent spin-bag quasiparticles (magnetic polarons).
We present results for the charge-density response function and for the optical
conductivity at zero temperature for different values of $t/J$.
They agree well with numerical results calculated by exact
diagonalization techniques. 
The density response function for intermediate and large momenta 
shows a broad continuum on energy scales of order of several $t$
whereas the optical conductivity for $\omega > 0$ is dominated by 
low energy excitations (at 1.5--2$J$).
We show that these weak-doping properties 
can be well understood by transitions between excited 
states of spin-bag quasiparticles.
\\
\end{abstract}

\widetext
\begin{multicols}{2}
\narrowtext


\section{Introduction}

The discovery of the high-temperature superconductivity has increased the 
interest in strongly correlated electronic systems. Many properties of
the superconducting cuprates are still not completely understood,
an effective theory capable of modeling the ground state and the excitations
of these materials remains an outstanding problem. 
Among the interesting features in the optical response of the cuprate 
materials are mid-infrared (MIR) structures at energies at 0.2-0.5 eV,
see e.g. \cite{ExpOptCon,Dagotto94}. Their origin has been discussed extensively
and is not yet completely clarified.

The one-band Hubbard model and the $t$-$J$ model are two candidates for
a description of the copper-oxide planes.
The purpose of this paper is an analytical study of the dynamical charge
response and the optical conductivity in the weakly doped $t$-$J$ model 
at finite energies and zero temperature.

Theoretical progress in this field is mostly based on numerical
techniques, especially on exact diagonalization 
methods\cite{ToHoMae95,EdOh95,EdOhMae95,EdWroOh96}
and Lanczos calculations\cite{JaPre95}. However, the small system sizes 
presently accessible to numerical methods leave many problems unresolved. 
So it is often not obvious whether the structures observed in the
spectra obtained from small-cluster calculations 
are finite-size effects or bulk
properties. Finite-size scaling is hard to perform for two-dimensional
systems because of the limited number of cluster sizes.
Small but finite momenta and hole concentrations cannot be investigated 
by numerics.

The spin response of doped antiferromagnets has also been studied in a number
of analytical papers but only few authors have investigated the optical 
conductivity and the charge density response function
\cite{Kotliar,GeZey95,Lee96,KhaHo96,ZeyKul96}.
The low-energy part of the optical conductivity has been found to fall off
slower than predicted by Drude theory; a pronounced MIR structure is visible
especially at zero or low temperature ($T \ll t,J$) and small doping 
($\delta < 10\%$).
Some of the analytical results indicate sharp peaks in the density response
at large momenta \cite{Kotliar,GeZey95}. A recent slave-boson approach
\cite{KhaHo96} yields one broad structure at high energy in this regime which 
is consistent with numerical studies. 
Besides the above mentioned fermionic models also boson models
have been studied, e.g. bosons (representing holes) coupled to a fluctuating
gauge field \cite{Lee96}. This coupling leads to an incoherent density 
fluctuation spectrum at finite temperatures.

The system we are interested in here is the 2D $t$-$J$ 
model\cite{Anderson87,Zhang88}:
\begin{equation}
H\, =\, - t \sum_{\langle ij\rangle \sigma}
      (\hat c^\dagger_{i\sigma} \hat c_{j\sigma} +
       \hat c^\dagger_{j\sigma} \hat c_{i\sigma})
    + J \sum_{\langle ij\rangle} \ ({\bf S}_i {\bf S}_j - {{n_i n_j} \over 4} )
\,.
\label{H_TJ}
\end{equation}
${\bf S}_i$ is the local electronic spin operator and $n_i$ the electron
number operator at site $i$. The symbol $\langle ij \rangle$ refers to
a summation over pairs of nearest neighbors.
At half filling the $t$-$J$ Hamiltonian reduces to the antiferromagnetic
Heisenberg model.
The electronic creation operators $\hat c_{i \sigma}^{\dagger}$
exclude double occupancies:
\begin{equation}
\hat c^{\dagger}_{i\sigma} = c^{\dagger}_{i\sigma} (1-n_{i,-\sigma})
\,.
\end{equation}
The $t$-$J$ model (\ref{H_TJ}) is by now believed to describe the relevant
low-energy degrees of freedom of the copper-oxide planes in the
high-$T_c$ materials. So we expect our results to coincide with 
experimental observations in the underdoped compounds at low energies 
only (1 eV and below).
Excitations at higher energies of course are not covered by the
$t$-$J$ Hamiltonian.

We study the time- and wave-vector dependent
charge-charge correlation function defined by
\begin{equation}
G_{\rho\rho}({\bf k},t) \,=\, 
 \langle\psi_0|\,\rho_{\bf k}^{el \dagger} \rho_{\bf k}^{el}(-t)\,
        |\psi_0\rangle
\end{equation}
The corresponding Laplace transform can be written as
\begin{eqnarray}
G_{\rho\rho}({\bf k},\omega) &=&
   \langle\psi_0|\,\rho_{\bf k}^{el \dagger} {1 \over {z-L}} \rho_{\bf k}^{el}\,
         |\psi_0\rangle\,,
\label{CCORR_DEF} \\
z&=&\omega+i\eta,\;\eta\rightarrow 0
\,.
\nonumber
\end{eqnarray}
Here,
$\rho_{\bf k}^{el} = \sum_{{\bf q},\sigma} 
c_{{\bf k+q},\sigma}^\dagger c_{{\bf q}\sigma}^{\phantom\dagger} =
\sum_{i\sigma} {\rm e}^{{\rm i}{\bf k}{\bf R}_i}
c_{i\sigma}^\dagger c_{i\sigma}^{\phantom\dagger}$
is the Fourier-transformed charge density operator.
$|\psi_0\rangle$ denotes the exact ground state of the system,
and $z$ is the complex frequency variable.
The Liouville operator $L$ is a superoperator defined by $L A = [H,A]_{-}$
for any operator $A$.
At zero temperature the optical conductivity $\sigma(\omega)$ 
for frequencies $\omega > 0$ is related to the
charge response function $G_{\rho\rho}({\bf k}, \omega)$ as follows:
\begin{equation}
{\rm Re}\,\sigma(\omega)\,=\, { {\rm e}^2 \over cN} \lim_{{\bf k} \rightarrow 0}
\frac {\omega\,{\rm Im}\,G_{\rho\rho}({\bf k}, \omega)} {{\bf k}^2}
\,
\label{OPTCON}
\end{equation}
where $N$ is the total particle number.
Eq. (\ref{OPTCON}) follows from the continuity relation
$L\rho_{\bf k}^{el} = {\bf k}\cdot{\bf j}_{\bf k}$ where
${\bf j}_{\bf k}$ is the Fourier-transformed charge current density operator. 

Due to strong correlations usual diagrammatic techniques based on Wick's
theorem can not be easily applied to solve the Hamiltonian (\ref{H_TJ}).
Neither the first nor the second part is
bilinear in fermion operators, and the
creation and annihiliation operators $\hat{c}_{i\sigma}^{\dagger}$
and $\hat{c}_{i\sigma}$ do not obey
simple anticommutation relations. For this reason, non-standard
analytical methods like variational wavefunctions, coupled-cluster
methods, slave-boson and slave-fermion
techniques or 1/N expansions have been applied to
the 2D $t$-$J$ model and related strongly correlated systems.

In the following, a projection technique\cite{Mori,Zwanzig,Forster}
approach based on the introduction of cumulants to evaluate charge density 
response functions is presented. 
We focus here on the finite-energy contributions:
Since we do not consider the diamagnetic 
part of the current we do not obtain results for the response at zero
energy, so the present calculation does not yield a Drude-like contribution 
to $\sigma(\omega)$.
In a recent letter \cite{VojBeck3} we have published first results 
for $G_{\rho\rho}({\bf k}, \omega)$ and $\sigma(\omega)$ using the present
method. Here we employ an improved set of projection variables and present
more details of the calculation.
At intermediate and large momenta we find for the density response function
a broad continuum of excitations on energy scales of several $t$.
The optical conductivity for finite $\omega$ 
is dominated by a small number of peaks at low
energies of order $J$.
The features and the different scaling behavior of these spectra can be
explained in terms of internal degrees of freedom of the spin-bag 
quasiparticles.

The paper is organized as follows: In Sec. II we briefly sketch
the cumulant method proposed in refs.~\cite{BeckFul88,BeckWonFul89,BeckBre90}
and show how to calculate dynamical correlation functions.
The description of the ground state of the weakly doped $t$-$J$ model within 
the cumulant formalism is subject of Sec. III. The employed ground state
wavefunction consists of independent hole quasiparticles (magnetic polarons)
moving on an antiferromagnetic background. 
The choice of appropriate dynamical variables for the projection technique 
is shown in Sec. IV. The variables are constructed from path operators
which form the hole quasiparticles.
In Sec. V we present results for the charge-charge correlation function and
the optical conductivity. They will be compared with results from numerical 
investigations found in the literature. 
A discussion of the results will close the paper.


\section{Cumulant method for correlation functions}

A recently introduced approach for calculating expectation values and dynamical
correlation functions\cite{BeckFul88,BeckWonFul89,BeckBre90} 
is based on the introduction of cumulants. Provided that the Hamiltonian
of the system can be split into an unperturbed part $H_0$ and 
a perturbation $H_1$, $H=H_0+H_1$, with eigenstates and
eigenvalues of $H_0$ known, this method uses the decomposition
\begin{equation}
e^{\ -\lambda H} \,=\,
e^{\ -\lambda (H_1+L_0)}\ e^{\ -\lambda H_0} \,.
\label{EXPH_DECOMP}
\end{equation}
This relation can be proved by comparing the equations of motion of either side with
respect to $\lambda$. $L_0$ is the Liouville operator corresponding to $H_0$,
defined by the relation $L_0 A = [H_0,A]_{-}$ for any operator $A$.
Let us denote the ground state of the unperturbed Hamiltonian $H_0$
by $|\phi_0\rangle$ and its energy by $\epsilon_0$
\begin{equation}
H_0 |\phi_0\rangle = \epsilon_0|\phi_0\rangle .
\end{equation}

Here we are interested in calculating dynamical correlation functions of 
operators $B_{\nu}$
\begin{equation}
G_{\nu\mu}(\omega) \,=\, \langle\psi_0|\,\delta B_{\nu}^+ {1 \over {z-L}}
                   \delta B_{\mu}^{\phantom +}\, |\psi_0\rangle
\label{ARBKORR}
\end{equation}
where we have introduced 
$\delta B_{\nu} = B_{\nu} - \langle\psi_0| B_{\nu} |\psi_0\rangle$.
Using (\ref{EXPH_DECOMP}) one can show\cite{BeckBre90} that these
correlation functions can be rewritten as
\begin{equation}
G_{\nu\mu}(\omega) \,=\, \langle\phi_0|\,\Omega^+\,B_{\nu}^+ \left( {1 \over {z-L}}
                   B_{\mu}^{\phantom +} \right)^{\cdot} \, \Omega\,
                   |\phi_0\rangle^c
\,.                   
\label{KUMKORR}
\end{equation}
The operator $\Omega$ has
similarity to the so-called wave operator (or Moeller operator known from
scattering theory). 
Within cumulants it transforms the ground state $|\phi_0\rangle$ of the
unperturbed system into the exact ground state $|\psi_0\rangle$ of $H$.
Explicitly it is given by
\begin{equation}
\Omega = 1 + \lim_{x \to 0} {1 \over {x-(L_0+H_1)}}H_1 \,.
\end{equation}
The brackets $\langle\phi_0|\,...\,|\phi_0\rangle^c$ denote cumulant 
expectation values formed with $|\phi_0\rangle$.
The dot $\cdot$ indicates that the quantity inside $(...)^{\cdot}$
has to be treated as a single entity in the cumulant formation.
For a detailed discussion of cumulants see e.g. Kubo\cite{Kubo}.

The relation (\ref{KUMKORR}) can be applied to either weakly or strongly
correlated systems because its use is independent of the operator
statistics , i.e., it is valid for fermions, bosons or spins.
Note that cumulants ensure size consistency for any subsequent 
approximations for the wave operator $\Omega$.

Using Mori-Zwanzig projection technique\cite{Mori,Zwanzig} one can derive
the following set of equations of motion for the dynamical correlation
functions $G_{\nu\mu}(\omega)$:
\begin{equation}
\sum_{\nu} \left( z\delta_{\eta\nu}-\omega_{\eta\nu}-\Sigma_{\eta\nu}(\omega) \right)
  \,G_{\nu\mu}(\omega)\,\,=\,\,\chi_{\eta\mu} \, .
\label{PROJ_GLSYS}
\end{equation}
$\chi_{\eta\nu}$, $\omega_{\eta\nu}$, and $\Sigma_{\eta\nu}$ are the static
correlation functions, frequency terms, and self-energies, respectively.
They are given by the following cumulant expressions:
\begin{eqnarray}
\chi_{\eta\nu} \,&=&\,
  \langle\phi_0|\,\Omega^\dagger\,B_{\eta}^\dagger B_{\nu}^{\phantom +}\,
                   \Omega\,|\phi_0\rangle^c \, ,
  \nonumber\\
\omega_{\eta\nu} \,&=&\,
  \sum_{\lambda}   \langle\phi_0|\,\Omega^\dagger\,B_{\eta}^\dagger
                   (LB_{\lambda}^{\phantom +})^{\cdot}\,
                   \Omega\,|\phi_0\rangle^c \,\chi_{\lambda\nu}^{-1} \, ,
\label{MATRIX_KUMDEF} \\
\Sigma_{\eta\nu}(\omega) \,&=&\, 
  \sum_{\lambda}   \langle\phi_0|\,\Omega^\dagger\,B_{\eta}^\dagger
                   \left( LQ {1 \over {z-QLQ}} L B_{\lambda}^{\phantom +} \right)^{\cdot}
                   \Omega\,|\phi_0\rangle^c 
\nonumber \\ &&\qquad\qquad\qquad\qquad\qquad\qquad\qquad\qquad \chi_{\lambda\nu}^{-1} \,.
\nonumber
\end{eqnarray}
$\chi_{\nu\mu}^{-1}$ is the inverse matrix of $\chi_{\nu\mu}$, 
and $Q$ is given by
\begin{equation}
Q = 1-P\, ,\quad
P = \sum_{\nu\mu} B_{\nu}^{\phantom +}\Omega|\phi_0\rangle^c\,\,
                   \chi_{\nu\mu}^{-1} \,\,
                   \langle\phi_0|\Omega^\dagger B_{\mu}^\dagger
\;.
\end{equation}
$P$ denotes a projection operator projecting onto the subspace of the Liouville
space spanned by the operators $B_{\nu}$, $Q$ projects onto the complementary
subspace. 

The described cumulant version of projection technique has a conceptual
advantage compared to standard projection technique.
Using projection technique the Laplace transform of a correlation function 
may be written as a continued fraction expansion (or a set of equations of
motion) describing the dynamics of the system. 
The continued fraction contains expectation values which are static
quantities. These expectation values have to be evaluated with the 
ground state of the interacting system. In standard projection technique 
this has to be done separately using mean-field methods or perturbation
theory. 
In contrast, in the cumulant approach static and dynamical aspects of the
system are treated along the same lines. For a further discussion
see \cite{BeckBre90}.


\section{Ground state wavefunction}

Now we turn to the $t$-$J$ model at weak doping.
The description of the ground state within the cumulant formalism 
used here has been shown to produce reasonable results for the 
doping dependence of the staggered magnetization and the spin-wave spectrum 
of the antiferromagnetic phase of the doped system\cite{VojBeck96}.

In the following $\delta$ denotes the hole concentration away from half filling.
Our system with $N$ lattice sites possesses $M=\delta N$ dopant holes.
The Hamiltonian is decomposed into $H_0$ and $H_1$ as follows:
\begin{eqnarray}
H_{0} &=& H_{\rm Ising} \,=\, J
  \sum_{<ij>}(S_{i}^{z}S_{j}^{z} - \frac{n_{i} n_{j}} {4} )
  \,+\, J (N-2M) \, , \nonumber \\
H_{1} &=& H_{t}\, +\, H_{\bot} \\
&=& -t \,\sum_{<ij>,\sigma}(\,\hat{c}_{i\sigma}^{+}\,
  \hat{c}_{j\sigma}\, \,+\, \,\hat{c}_{j\sigma}^{+}\,\hat{c}_{i\sigma}\,)
\nonumber \\
&&  \,+\, \frac{J} {2}\,\sum_{<ij>}(\, S_{i}^{-}S_{j}^{+}\,+\, S_{i}^{+}S_{j}^{-}
  \, )\, \nonumber.
\end{eqnarray}
The ground state $|\phi_0\rangle$ of the unperturbed Hamiltonian $H_0$ is an
antiferromagnetically ordered N\'{e}el state with $M$ holes. The holes have
fixed momenta ${\bf k}_m$ and are located on the sublattice $\sigma_m$
($\sigma_m = \uparrow,\downarrow$)
\begin{eqnarray}
|\phi_0\rangle\, &=&\,
  \hat{c}_{{\bf k}_{1}\sigma_{1}}\, \ldots\,\hat{c}_{{\bf k}_{M}\sigma_{M}}
  \, |\phi_{N{\mathaccent 19 e}el}\rangle
\label{GSDEF3} \\
&=&\, \prod_{m=1}^{M} \left( \frac{1} {\sqrt{N/2}}
  \sum_{i_{m}\in\sigma_m}\,
  {\rm e}^{i\,{\bf k}_{m}{\bf R}_{i_m}}
  \hat{c}_{i_m\sigma_m}\right) |\phi_{N{\mathaccent 19 e}el}\rangle \, .
\nonumber
\end{eqnarray}
In $H_0$ we have shifted the energy level so that 
$\langle\phi_0|H_0|\phi_0\rangle = 0$.

Within the cumulant method we employ an exponential ansatz\cite{SchorkFul92} 
for the wave operator $\Omega$, $\Omega = {\rm e}^S$. Here, the operator
$S$ contains the effect of the perturbation $H_1$ onto the unperturbed
ground state $|\phi_0\rangle$. We have to consider two parts of $H_1$,
the spin-flip term $H_\perp$ and the hopping term $H_t$.

The hole motion processes induced by $H_t$ are described by path 
operators\cite{Nagaoka,BrinkRice,Trugman88,ShrSig88}
which leads to the picture of spin-bag quasiparticles\cite{Schrieffer}.
Here we define path concatenation operators $A_{n,\xi}(i)$ where $i$ denotes
a lattice site, $n$ the path length, and $\xi$ the individual path shape:
$A_{n,\xi}(i)$ operating on the N\'{e}el state with one hole at site $i$,
$\hat c_{i\uparrow} |\phi_{N{\mathaccent 19 e}el}\rangle$, moves the hole
$n$ steps away and creates a path or string $\xi$ of $n$ spin defects
attached to the transferred hole. For $n=1$ there are 4 different path
shapes (Fig.~1), for $n=2$ there are 12 and so on.
Explicitly, the operators $A_{n,\xi}(i)$ for sites $i$ on the
$\uparrow$-sublattice are defined by
\begin{eqnarray}
A_{1,\xi}(i)\,\, &=&\,\, \sum_{j}\hat{c}_{j\downarrow}
        \hat{c}_{i\downarrow}^{+}\, R_{ji}^\xi\, ,\nonumber\\
A_{2,\xi}(i)\,\, &=&\,\, \sum_{jl}
        \hat{c}_{l\uparrow}S_{j}^{+}\hat{c}_{i\downarrow}^{+}
        \, R_{lji}^\xi\, ,\quad
        (i,l\in\uparrow,\, j,m\in\downarrow) \\
A_{3,\xi}(i)\,\, &=&\,\, \sum_{jlm}
        \hat{c}_{m\downarrow}S_{l}^{-}S_{j}^{+}\hat{c}_{i\downarrow}^{+}
        \, R_{mlji}^\xi \, ,\nonumber\\
\ldots\quad&\quad&\nonumber
\end{eqnarray}
The operators $A_{n,\xi}$ for sites $i$ on the
'down' sublattice are defined analogously with all spins reversed.
The matrices $R_{i_n ... i_1}^\xi$ allow the hole to jump along a path
of shape $\xi$:
\begin{eqnarray}
R_{i_n ... i_1}^\xi &=& \left\{
   \begin{array}{rl}
     1&\, i_1, ..., i_n\,\, {\rm connected\,by\,pathshape}\,\xi\\
     0&\, {\rm otherwise}\end{array}\right. .
\end{eqnarray}

\end{multicols}
\widetext

\begin{figure}
\epsfxsize=14cm
\centerline{\epsffile{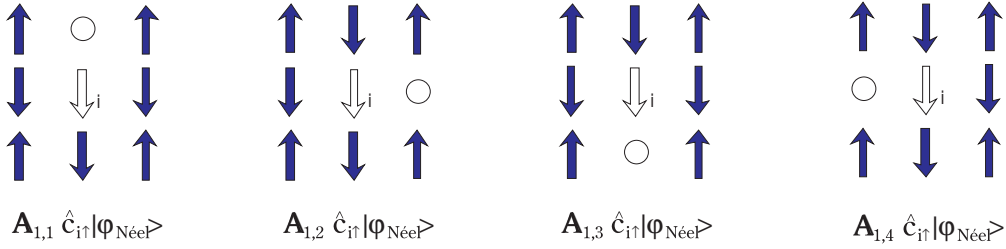}}
\caption{Path shapes of length 1 created by the operators $A_{1,\xi}$
         acting on a hole at site $i$ in the $\uparrow$-sublattice.}
\end{figure}

\begin{multicols}{2}
\narrowtext

Having defined the excitation operators $A_{n,\xi}(i)$ 
the wave operator $\Omega$ used for the cumulant formalism takes the form
\begin{eqnarray}
\Omega \:&=&\: \exp \left(
  \sum_{n=1}^{n_{\rm max}}\sum_{\xi=1}^{m_n} \lambda_{n,\xi} A_{n,\xi}\, \right) \,,
\label{STOEROPDOT2}
\\
A_{n,\xi}\:&=&\:\sum_i A_{n,\xi}(i)
\nonumber
\end{eqnarray}
with parameters $\lambda_{n,\xi}$ yet unknown.
Note the additional definiton 
$A_0(i) = {1 \over 2} \sum_\sigma 
\hat c_{i\sigma}^{\phantom\dagger} \hat c_{i\sigma}^\dagger=
(1-n_{i\uparrow})(1-n_{i\downarrow})$
which is the only "path" with zero length ($\xi=1$). This operator
is a projection operator on the empty state at site $i$.
The path operators $A_{n,\xi}$ commute with each other because they only
contain spin-flip operators destroying N\'{e}el order. 
Applying the operator $\Omega$ (\ref{STOEROPDOT2}) to the unperturbed
ground state (\ref{GSDEF3}) adds to each hole a cloud of spin defects 
leading to a spin bag or magnetic polaron \cite{ShrSig88,Schrieffer}.
In the ansatz (\ref{STOEROPDOT2}) we use separate coefficients
$\lambda_{n,\xi}$ for all individual paths $A_{n,\xi}$. This is an
extension compared to the ansatz of ref. \cite{VojBeck96} 
where all paths with the same length $n$ had been given the same weight. 
The additional degrees of freedom improve the ground-state wavefunction 
for non-zero hole momentum taking into account that
the quasiparticle state is not exactly s-like, see also
Sec. V C.

The transverse part $H_\perp$ of the magnetic exchange creates or destroys 
pairs of spin defects and therefore can change the length of a given path of
spin defects by 2. This process leads to a coherent motion of the hole
quasiparticle through the lattice \cite{BrinkRice,Trugman88}.

Additional spin fluctuations in the antiferromagnetic background will be
neglected here. The results show
that this approximation gives already reliable results for the charge 
density response of the system because the charge dynamics is mainly 
determined by the motion of the hole quasiparticles. 
However, for a description of magnetic properties or the spin response 
of the system these ground-state
spin fluctuations would be essential, see e.g. \cite{VojBeck96}.

The coefficients $\lambda_{n,\xi}$ of the ansatz (\ref{STOEROPDOT2}) are 
determined by the condition for $\Omega |\phi_0\rangle$ to be an eigenstate
of $H$.
Following ref.\cite{SchorkFul92} one arrives at a non-linear set of equations
for the ground-state energy $E_0$ and the coefficients $\lambda_{n,\xi}$:
\begin{eqnarray}
E_{0}\,=\, \langle\phi_0| H \Omega |\phi_0\rangle^c \, , \quad
0    \,=\, \langle\phi_0| A_{n,\xi}^\dagger\, H \, \Omega\, |\phi_0\rangle^c
\label{GLSYSPFADSPINW}
\end{eqnarray}
with $\Omega$ given by (\ref{STOEROPDOT2}).
The cumulant expectation values have to be taken with respect to the unperturbed
ground state (\ref{GSDEF3}). 
The set of equations (\ref{GLSYSPFADSPINW}) together with the additional 
approximation of independent hole quasiparticles leads to a generalized
eigenvalue problem. For details see ref.\cite{VojBeck96}. 
The total ground-state
energy depends on the initial momenta of the hole quasiparticles ${\bf k}_m$
in (\ref{GSDEF3}). The energy minimum for the one-hole problem is located at
${\bf k} = (\pm\pi/2,\pm\pi/2)$ which leads to a hole-pocket Fermi 
surface. Further properties of the one-hole spectrum are discussed 
in Sec. V C.
The used picture of
independent quasiparticles is appropriate for small hole concentrations,
i.e., for a dilute gas of holes moving in an antiferromagnetic background.


\section{Dynamical variables for charge correlations}

To calculate dynamical charge correlation functions we employ
the cumulant version of Mori-Zwanzig projection technique as described
in Sec.~II. 
At first we have to choose
a set of relevant operators $B_{\nu}$. Here we are going to neglect the
self-energy terms $\Sigma_{\nu\mu}$ from (\ref{MATRIX_KUMDEF}).
This can be done using a sufficiently
large set of operators $B_{\nu}$ to cover the charge dynamics we are 
interested in.

One variable to be included is the charge density operator itself.
Note that the particle density 
$\rho_{\bf k}^{el} = \sum_{i\sigma} e^{{\rm i} {\bf k} {\bf R}_i}
{\hat c}_{i\sigma}^\dagger {\hat c}_{i\sigma}^{\phantom\dagger}$
and the hole density 
$\rho_{\bf k}^{hole} = {1 \over 2} \sum_{i\sigma} e^{{\rm i} {\bf k} {\bf R}_i}
{\hat c}_{i\sigma}^{\phantom\dagger} {\hat c}_{i\sigma}^\dagger
= \sum_{i\sigma} e^{{\rm i} {\bf k} {\bf R}_i} A_0(i)$
are equivalent quantities according to
\begin{eqnarray}
\rho_{\bf k}^{hole}+\rho_{\bf k}^{el} &=&
  \sum_{i\sigma} e^{{\rm i} {\bf k} {\bf R}_i}
  ( {1 \over 2}
    {\hat c}_{i\sigma}^{\phantom\dagger} {\hat c}_{i\sigma}^\dagger +
    {\hat c}_{i\sigma}^\dagger {\hat c}_{i\sigma}^{\phantom\dagger} )
\nonumber \\
 &=&  N\,\delta_{{\bf k},0}\,.
\end{eqnarray}
Note that
$\sum_\sigma \,\, ( {1 \over 2}
    {\hat c}_{i\sigma}^{\phantom\dagger} {\hat c}_{i\sigma}^\dagger \,+\,
    {\hat c}_{i\sigma}^\dagger {\hat c}_{i\sigma}^{\phantom\dagger} )$
is the projector on the empty and singly occupied states at site $i$.

As first variable we therefore choose $\rho_{\bf k}^{hole}$. 
Additional dynamical variables 
$B_\nu$ can be deduced from the action
the Liouville operator to the first variable $\rho_{\bf k}^{hole}$,
i.e. from $L\,\rho_{\bf k}^{hole}$, $L (L\,\rho_{\bf k}^{hole})$ etc. 
All variables should provide particle number conservation because
$H$ and the density operator do not change the total number of particles.

The hopping part of $L$ creates strings of spin defects
when applied to $\rho_{\bf k}^{hole}$, i.e., $L_t^n\,\rho_{\bf k}^{hole}$ 
is a sum of path concatenation operators up to length $n$. 
The spin-flip part $L_\perp$ creates or destroy pairs of
spin defects. It  
changes the length
of a spin-defect path by 2. Additional processes connecting paths
of different holes, i.e. hole-hole interactions, will be neglected
consistently with the ground state evaluation procedure.
Assuming that the charge dynamics is mainly carried by the spin-bag
quasiparticles we define a set of variables which contains the
processes described above:
\begin{eqnarray}
B_{n,\xi,{\bf\Delta}}({\bf k})
&=&
  \sum_{i,\sigma} e^{{\rm i} {\bf k} {\bf R}_i} \,\,
  (A_{n,\xi}(i+\Delta){\hat c}_{i+\Delta,\sigma})\,
  (A_0(i){\hat c}_{i\sigma})^\dagger
\nonumber\\
&=&
  \sum_{i,\sigma} e^{{\rm i} {\bf k} {\bf R}_i} \,\,
  B_{n,\xi,{\bf\Delta}}(i)
\label{PROJSET}
\end{eqnarray}
where $A_{n,\xi} (i)$ is a path operator of length $n$ with the path shape
$\xi$ acting on a hole at site $i$, as defined in the previous section.
The indices $n,\xi,\bf\Delta$ replace the general index $\nu$ of the dynamical
variables $B_{\nu}$ in Eqs. (\ref{ARBKORR},\ref{KUMKORR}).
The operator $B_{n,\xi,{\bf\Delta}}$ couples to a hole at site $i$ destroying it,
creates a hole at site ${\bf R}_i+{\bf R}_\Delta$ and adds the
path $A_{n,\xi}$. 
Here, ${\bf R}_\Delta$ is a vector connecting two sites on the same sublattice,
$i+\Delta$ is a short-hand notation for ${\bf R}_i+{\bf R}_\Delta$.
As an example, Fig.~2 shows the effect of the
variable $B_{2,4,0}$ applied to a path state
$A_{3,1} {\hat c}_{i\uparrow} |\phi_{N{\mathaccent 19 e}el}\rangle$.
Variables with ${\bf\Delta}\neq 0$ arise from the application
of $H_\perp$ at sites at the beginning of a spin-defect path. Such a process
shortens the path by 2 and moves the initial site of the quasiparticle 
(the path starting point) two sites away ($|{\bf R}_\Delta| = 2$).
The operators $B_{n,\xi,\bf\Delta}$ provide a coupling between the
ground state and excited states of the spin-bag quasiparticles,
see Sec. V C.
So the relevant part of the Liouville space spanned by the operators 
$B_{n,\xi,\bf\Delta}$ contains all individual path states of the holes.
Note that the first of these operators is the hole density operator itself:
\begin{equation}
B_{0,1,0}({\bf k}) \,=\, \rho_{\bf k}^{hole} \,=\,
  {1 \over 2}
  \sum_{i\sigma} {\rm e}^{{\rm i} {\bf k} {\bf R}_i} \,\,
  {\hat c}_{i\sigma}^{\phantom +} {\hat c}_{i\sigma}^\dagger
\end{equation}
The quantity we are interested in
is therefore the diagonal correlation function $G_{\nu\nu}({\bf k},z)$
with $\nu=(0,1,0)$.

Additional projection variables could be found by applying the spin-flip term
$H_\perp$ to $B_{n,\xi,\bf\Delta}$ at sites along the path. 
Those variables would provide a coupling to states with spin fluctuations 
separated from the spin-bag quasiparticles.
Such processes would describe scattering of the quasiparticle
states with spin waves, they will be neglected here.

The above set of dynamical variables is an extension of the one used in
previous calculations \cite{VojBeck3}. 
There we had included only variables with ${\bf\Delta}=0$. 
The new variables with ${\bf\Delta}\neq 0$ provide additional
degrees of freedom being useful if more than one hole is present in the 
system. 
This can be easily understood from a comparison with the density response 
of a system of free fermions:
The correct excitations are obtained there if one splits the density
$\rho_{\bf k} = \sum_{\bf q} c_{\bf k+q}^\dagger c_{\bf q} =
\sum_i e^{{\rm i} {\bf k R}_i} c_i^\dagger c_i^{\phantom\dagger}$ into 
a set of dynamical variables $\{c_{\bf k+q}^\dagger c_{\bf q}\}$ where
$\bf q$ are the momenta of the particles present in the ground state. 
An equivalent set of variables can be obtained from linear combinations
of the $\{c_{\bf k+q}^\dagger c_{\bf q}\}$ leading to 
$\{\sum_i e^{{\rm i} {\bf k R}_i} c_{i+\Delta}^\dagger c_i^{\phantom\dagger}\,\}$
which correspond to the $B_{n,\xi,\bf\Delta}$ of the present
calculation.

\begin{figure}
\epsfxsize=7cm
\centerline{\epsffile{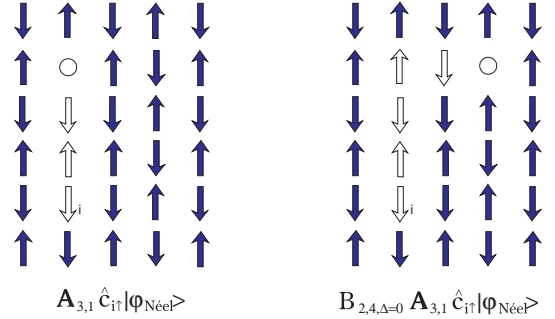}}
\caption{Effect of $B_{2,4,0}$ on a path of length 3 created by $A_{3,1}$.}
\end{figure}

Having chosen the set of projection variables (\ref{PROJSET}) we can 
consider the equations of motion (\ref{PROJ_GLSYS}) for the
correlation functions of the operators $B_{n,\xi,\bf\Delta}$.
The relevant terms for the dynamics are the static correlation functions
and the frequency terms given in (\ref{MATRIX_KUMDEF}).
The arising cumulant expectation values can be transformed to normal 
expectation values according to appendix A. Calculating the matrix elements
we have neglected hole-hole interaction processes and geometry effects such
as spiral paths. Within these approximations the path operators form an
orthogonal basis set,
\begin{equation}
  \langle\phi_0| A_{n,\xi}^\dagger A_{m,\eta}^{\phantom\dagger} |\phi_0\rangle
  \,\,=\,\, M\,\delta_{nm}\delta_{\xi\eta} 
\,,
\end{equation}
where $M$ is the number of holes present in the system. 
The static matrix $\chi$
from (\ref{MATRIX_KUMDEF}) therefore becomes proportional to a unity matrix:
\begin{equation}
  \langle\phi_0| \Omega^\dagger\,
  B_{n,\xi,{\bf\Delta}}^\dagger \,
  B_{m,\eta,{\bf\Theta}}^{\phantom\dagger} \,
  \Omega |\phi_0\rangle^c
  \,\,=\,\, M\,\delta_{nm}\delta_{\xi\eta}\delta_{\bf\Delta\Theta}
\,.
\end{equation}
The elements of the frequency matrix from (\ref{MATRIX_KUMDEF}) 
contain the hole motion processes. 
To be short, here we only state the results for one hopping process:
\begin{eqnarray}
  \langle\phi_0| \Omega^\dagger\,
  B_{n,\xi,{\bf\Delta}}^\dagger 
  (L_t \, B_{m,\eta,{\bf\Theta}}^{\phantom\dagger} )^\cdot \,
  \Omega |\phi_0\rangle^c \nonumber\\
  = t\, M\,(\delta_{n\xi,m\eta+1}\,+\,\delta_{n\xi+1,m\eta})
                    \delta_{\bf\Delta\Theta}
\end{eqnarray}
where $\delta_{n\xi,m\eta+1}$ is 1 if the path $A_{n\xi}$ is obtained
from $A_{m\eta}$ by one further hopping process, otherwise it is zero.
For more details concerning the matrix elements we refer to a recent
publication\cite{BEW}.


\section{Results}

In the numerical calculations we have included paths up to length 5 which gives
a set of 485 path variables $A_{n,\xi}$. 
The system has a few holes ($\delta \ll 1$) in each of the four hole pockets 
(at momenta $(\pm\pi/2,\pm\pi/2)$). It can be shown that in this case
four different vectors $\bf\Delta$ in the set of dynamical variables 
are sufficient to describe the charge dynamics at low doping. Note that
this approximation does not take into account changes of the hole Fermi 
surface with doping, so our treatment is strictly valid only in the limit of
$\delta \rightarrow 0$.
These Fermi surface changes would be important for a good description of 
the low-energy intraband excitations of the hole quasiparticles at
non-zero doping (see below).

We have tested
maximum path lengths of 2, 3, 4, and 5. Beyond a path length of 2 we obtained
no essential differences in the spectra except of richer structures with an 
increasing number of projection variables. 
Thus we expect only minor changes when including longer paths, see Sec. V C.

\subsection {Charge correlation function}

Results for the charge response function $G_{\rho\rho}({\bf k},\omega)$ at
$t/J$ = 1, 2.5, and 5 are shown in Fig. 3.
Note that we obtain discrete spectra because we have neglected all 
self-energies.
In the figures we have introduced Lorentzians with a small artificial 
linewidth.
The main doping dependence of the spectra is found in the
intensity which is proportional to the hole concentration.
This follows from the assumption of independent hole motion and a 
doping-independent spin background (rigid-band picture). 
Within these approximations
all matrix-elements in (\ref{MATRIX_KUMDEF}) are essentially
proportional to the hole concentration $\delta$. From (\ref{PROJ_GLSYS})
it is seen that therefore all correlation functions $G_{\nu\mu}$ are
proportional to $\delta$.

For smaller momentum the spectral weight is mainly
concentrated in a peak near $\omega=0$.
With increasing momentum we find a transfer of spectral weight 
to higher energies. The figures for different values of $t/J$ show that the
broad structures observed in the spectra for large momenta scale with $t$,
e.g., the maximum spectral weight in the charge-response function
at $(\pi,\pi)$ remains at energies of about $6t$ independent of 
$t/J$.
This scaling behavior will be discussed in the following sections.

Next we compare our data with exact diagonalization (ED) results. 
Due to the small cluster sizes no numerical spectra are available for small 
hole concentrations.
Fig. 4. shows a comparison of our data for $t/J = 2.5$ with ED
data for $\delta = 25\%$ taken from ref.\cite{ToHoMae95}. 
It is remarkable that even for this large hole concentration we observe 
a qualitatively good agreement with our spectra: 
Both results show the weight transfer to higher energies with increasing
momentum and a broad continuum of excitations at large momenta.
Differences may be either due to the small number of sites in 
the numerical calculations (which leads to an energy gap in
all spectra and other finite-size effects) 
or due to the neglection of relaxation processes 
(which would produce finite linewidths) and Fermi-surface
effects (which would affect the low-energy spectrum for non-zero
doping) in the present calculation.

\end{multicols}
\widetext

\begin{figure}
\epsfxsize=14cm
\centerline{\epsffile{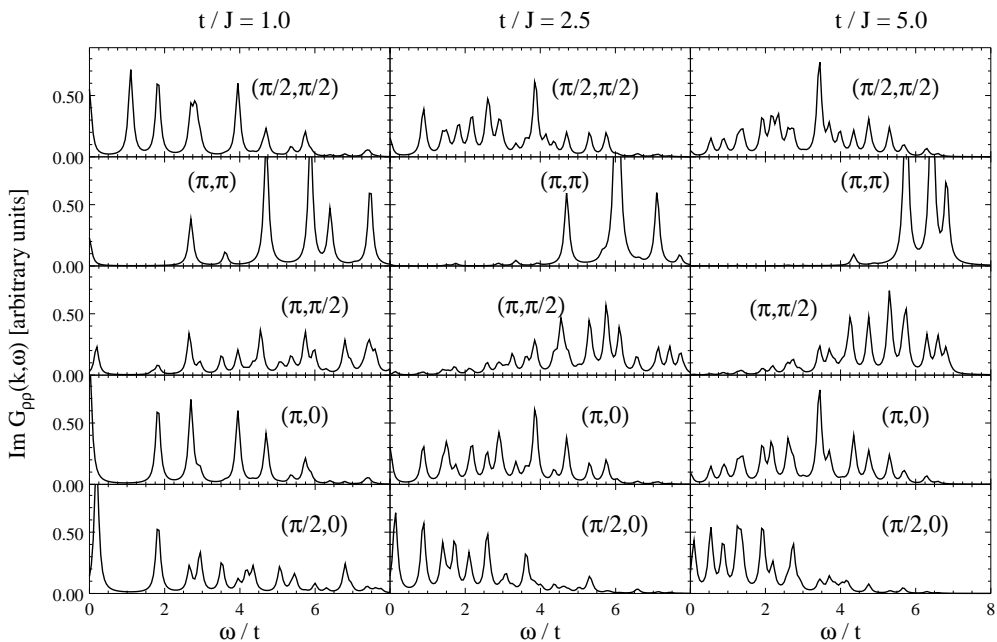}}
\caption{Charge response functions obtained from the 
present calculation for very small $\delta$,
different momentum transfers and 
parameter values $t/J$=1, 2.5, and 5.
The peaks are plotted with Lorentzians using an artificial 
linewidth of 0.1 $t$.}
\end{figure}

\begin{figure}
\epsfxsize=11cm
\centerline{\epsffile{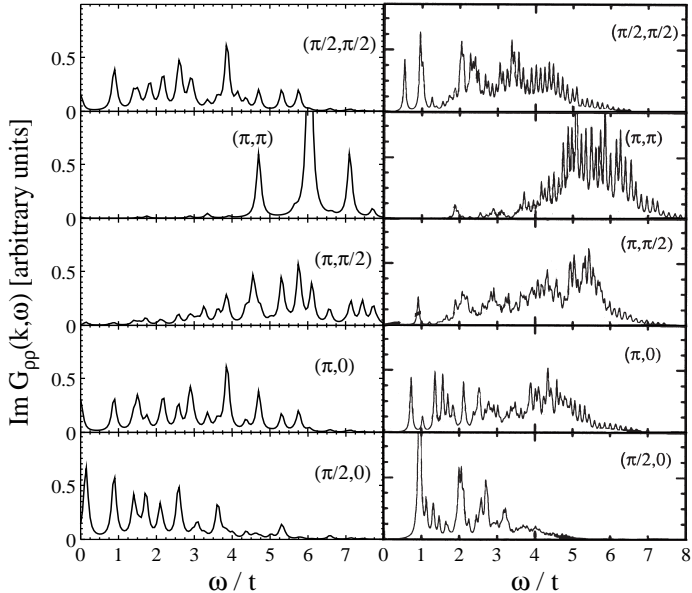}}
\caption{Left panel: charge response function from Fig. 3 for $t/J$ = 2.5.
Right panel: Exact diagonalization data from ref. \protect\cite{ToHoMae95}
for comparison. $t/J = 2.5$. The numerical
data have been calculated for a 4 x 4 periodic cluster and a hole
concentration of $\delta = 25\%$.}
\end{figure}

\begin{multicols}{2}
\narrowtext

\subsection{Optical conductivity}

Using Eq. (\ref{OPTCON}) one can deduce the optical conductivity from
the calculated charge response function.
Results for ${\rm Re}\,\sigma(\omega)$ are displayed in Fig.~5.
We find a few peaks at low energy, the positions of these main peaks 
scale with $J$. 
These features at 1.5 -- 2 $J$ were
also found in numerical studies of the $t$-$J$ and Hubbard models.
They are supposed to coincide with the MIR structures observed in
optical spectra of high-$T_c$ superconductors being located at 
0.2--0.5 eV (for $t\approx$ 0.5 eV, $J\approx$ 0.15--0.2 eV),
see e.g.\cite{Dagotto94}.

Within the present calculation we do not obtain a Drude-like contribution
to $\sigma(\omega)$ since eq.
(\ref{OPTCON}) is valid for non-zero frequency only, i.e., it does
not include the diamagnetic part of the current.
(Due to the neglection of the self-energies which excludes scattering
processes of the quasiparticle states and the assumption of
independent hole motion the Drude contribution would have the form
$D\,\delta(\omega)$.)
So the present calculation cannot account for analytical features as, 
e.g., the drop-off of the Drude peak of finite $\omega$.

Comparing our results with numerical data (at higher doping) we
again observe reasonable agreement e.g. with the ED results
from ref.\cite{EdWroOh96} for $\delta = 12.5\%$.

\end{multicols}
\widetext

\begin{figure}
\epsfxsize=14cm
\centerline{\epsffile{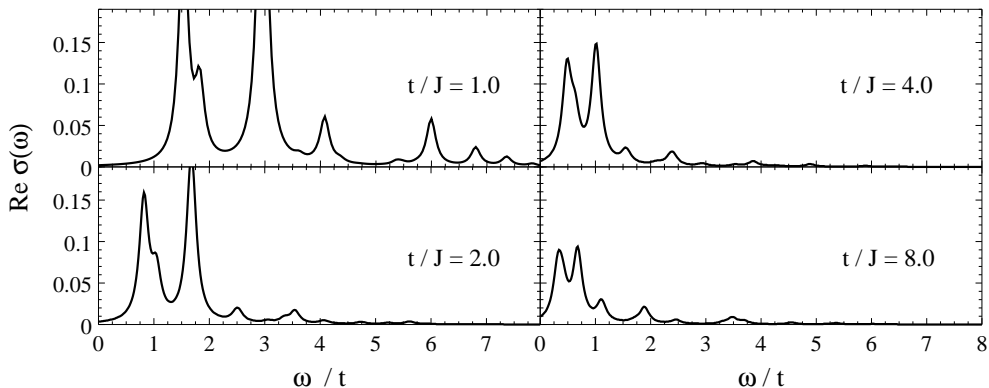}}
\caption{Optical conductivity for small hole concentration and different
$t/J$. The spectra have been calculated from the charge-charge correlation
function using (\protect\ref{OPTCON}).}
\end{figure}

\begin{multicols}{2}
\narrowtext

\subsection{Relation to the one-particle spectrum}

For the interpretation of the calculated spectra one can consider
the one-hole spectrum of the Hamiltonian within our approximation.
From the diagonalization of the one-hole problem in the subspace
of path operators $A_{n,\xi}$ one obtains several bands for the
spin-bag quasiparticles.
As is well known, the lowest band has a
minimum at $(\pm\pi/2,\pm\pi/2)$ and a bandwidth of about 2$J$.
It corresponds to a quasiparticle with s-like symmetry. 
The spin-bag states in the higher bands have nodes in the coefficients 
$\lambda_{n,\xi}$ (compare (\ref{STOEROPDOT2}) ). 
Note, however, that a classification according 
to angular momentum (p, d, f and so on) is not appropriate 
at least for longer paths because the
path states $A_{n,\xi}$ do not obey simple rotational symmetry.
The total number of one-particle states is of course equal to the number
of considered path variables. Increasing the maximum path length $n_{max}$ adds
essentially many non-s-like states to the spectrum which have energies
between the s-like states. So our results do not depend strongly
on the maximum path length beyond $n_{max} > 2$.

The low-energy peak at small, but finite momenta is caused
by excitations within the lowest quasiparticle band.
These excitations are treated correctly here only for the
limit $\delta \rightarrow 0$ since we have neglected changes
of the Fermi surface with doping.
The next structures at low energies arise from transitions between the first and second/third
quasiparticle band, i.e., from the s-like groundstate to p-like states.
These structures are especially visible at small momentum transfer, i.e., 
in the optical conductivity, see the discussion below. 
The high-energy part of the spectra corresponds to excitations to 
higher bands.
So our calculation supports the discussion given in ref.\cite{EdWroOh96} 
where only the two low-lying states have been considered in a simplified 
analytical calculation for the main peak found in the optical spectrum.
We want to emphasize that taking into account all string states
(up to a truncation length) as done in the present work reproduces not
only this main peak but the incoherent continuum at energies
up to $8t$ also found in numerical work.
So we have mapped the charge excitations for the $t$-$J$ model (\ref{H_TJ}) 
at small hole concentration to transitions between
bands of non-interacting spin-bag quasiparticles. 
(Not taken into account are changes of the Fermi surface of these quasiparticles
with doping and scattering of the quasiparticle states.)

The different scaling behavior of $G_{\rho\rho}(\bf k,\omega)$ (for
large $\bf k$) and $\sigma(\omega)$ with $t/J$ arises from the internal
structure of the spin-bag quasiparticles. The width of the lowest
band as well as low excitation energies scale primarily with $J$, i.e.,
the excitation energies can be written as $a t + b J$ with 
$b/a \gg t/J \approx 2...5$. 
Therefore the excitation from the ground state to the first excited 
state has an energy of order $J$ as seen in the optical conductivity.
At large $\bf k$ the energies of the structures in the density response
scale with $t$. This can be understood from the following qualitative
arguments: 
At large $\bf k$ the operator $\rho_{\bf k}^{hole}$ couples the 
ground state to higher excited states of the quasiparticle. This coupling
depends on $t/J$ since the quasiparticle size changes with $t/J$. 
(Mapping the localized hole problem onto its continuum version\cite{Kane} 
one finds that the path coefficients decrease exponentially with a 
length scale proportional to $(t/J)^{1/3}$.)
Thus the spatial distance of the real-space nodes in the coefficients 
$\lambda_{n,\xi}$ increases with $t/J$ (for fixed $n,\xi$),
in other words, for a fixed spatial node distance the excitation number
$n$ increases with $t/J$.
Therefore with increasing $t/J$ the operator $\rho_{\bf k}^{hole}$ 
(for fixed $\bf k$) couples to higher quasiparticle states. 
This results in structures scaling
with $t$ in the density response function for large $\bf k$.

It is worth noting that the non-s-like states of the quasiparticle
play an important role for the charge dynamics considered here. 
Neglecting them, i.e. considering all paths with the same length as ONE
dynamical variable, would be a bad approximation. 
However, for the one-particle dynamics these states do nearly not contribute
to the spectrum: If one calculates the one-hole spectral function
in the subspace of path operators then only the s-like states carry 
considerable spectral weight. The reason for this behavior lies in the
symmetry of the Hamiltonian.


\section{Conclusion}

In the present paper we have studied the charge dynamics in
weakly doped antiferromagnets described by the $t$-$J$ model 
at zero temperature.
Our ansatz for the ground-state wave-function\cite{VojBeck96}
includes mobile hole quasiparticles. Background spin fluctuations have been
neglected. Thus the ground state has antiferromagnetic long-range
order independent of hole doping.
We have used this ground state together with a cumulant version 
of Mori-Zwanzig projection technique to calculate dynamical charge
correlation functions.

We find structures at lower energies scaling with $J$ for zero and small
momentum
and a transfer of spectral weight to higher energies with increasing 
momentum. At intermediate and large momenta the density response spectrum consists
of a broad continuum at energies of order $t$.
The calculations do not cover a Fermi-surface related low-energy response
(including a Drude-like peak in the optical conductivity).
Our results for the density response function as well as the optical 
conductivity are in reasonable agreement with recent numerical
data obtained by exact diagonalizations\cite{ToHoMae95,EdWroOh96,KhaHo96}
(see Figs. 3-5).
However, these numerical calculations have been done for large hole concentrations
(e.g., 25\%) where only short-range magnetic order is present.
In contrast, our calculations are based on magnetic long-range order and
neglect background spin fluctuations.
Thus we conclude that the fact whether the quasiparticles move in a long-range
ordered background or not does not have an essential influence on the
charge response of the system at higher energies.
The dynamics is determined by the local antiferromagnetic order in the 
vicinity of the hole quasiparticles, i.e., the magnetic correlation length 
has to be of the order of the quasiparticle size. More precisely,
our treatment is valid with antiferromagnetic order on length scales of the
longest path variables included in the calculation (here 5 lattice constants).
Note here that the magnetic correlation length in the high-$T_c$ materials
is about 5 lattice constants for $\delta = 12.5\%$ and about 2.5 lattice 
constants for $\delta = 25\%$ differing somewhat for different 
compounds. (Numerical simulations for the $t$-$J$
and related models yield similiar values although precise data cannot
be obtained because of the restricted system sizes.)

A remaining task would be to include doping dependent background spin
fluctuations in our ansatz for the wave operator $\Omega$ as done in 
the static calculations of ref.\cite{VojBeck96}.
The scattering of quasiparticles at spin fluctuations would cause a decay
of the excited quasiparticle states and provide lifetime broadening of 
the lines in the response functions. From this we do not expect drastic 
changes in the response functions for larger energies.
A second possible improvement concerns the inclusion of Fermi-surface
related response at low energies, work along this line is in progress.

\appendix

\section{Evaluation of cumulants}

In this appendix we show how to evaluate cumulants with an exponential
ansatz for the wave operator $\Omega$.
The basic relation is
\begin{equation}
\langle\phi|\,{\rm e}^{S^+}\,\prod_i A_i^{n_i}\,{\rm e}^S\,|\phi\rangle^c\, = \,
{\langle\psi|\prod_i A_i^{n_i}|\psi\rangle^c }
\label{KUMREL1}
\end{equation}
with $A_i$ being arbitrary operators.
Note that on the l.h.s. the operators $S^+$, $S$ and $A_i$ are subject to
cumulant ordering whereas on the r.h.s. only the $A_i$ operators are
cumulant entities. However, the cumulants on the r.h.s. are formed
with the new wavefunction $|\psi\rangle = {\rm e}^S |\phi\rangle$.

Eq. (\ref{KUMREL1}) can be proven either by integrating infinitesimal transformations
$(1+S/N)$ and using properties of cumulants\cite{Kladko}
or by explicitly using the definition of cumulant expectation values.
Here we demonstrate the second way. We start from the definition of
cumulant expectation values\cite{Kubo} for
a product of arbitrary operators $A_i$ and an arbitrary state $|\phi\rangle$:
\begin{eqnarray}
\langle\phi|\prod_i A_i^{n_i}|\phi\rangle^c\, = \, \nonumber\\
\left(\prod_i \left({\partial \over \partial\lambda_i} \right)^{n_i} \right)
\ln\langle\phi|\prod_i {\rm e}^{\lambda_i A_i}|\phi\rangle\,
|_{\lambda_i=0\,\forall\,i}\,.
\end{eqnarray}
We consider the following expression:
\begin{eqnarray}
&&\langle\phi| {\rm e}^{\alpha S^+} \prod_i A_i^{n_i} \, {\rm e}^{\beta S} |\phi\rangle^c
  \nonumber \\
&&\, = \,
\sum_{n=0}^\infty \sum_{m=0}^\infty {\alpha^n \over {n!}} {\beta^m \over {m!}}
\langle\phi| {S^+}^n \prod_i A_i^{n_i} S^m |\phi\rangle^c \nonumber\\
&&\, = \,
\sum_{n=0}^\infty \sum_{m=0}^\infty {\alpha^n \over {n!}} {\beta^m \over {m!}} \,\,
\left({\partial \over \partial\xi} \right)^n
\left({\partial \over \partial\eta}\right)^m
\nonumber\\
&&\left[
\left(\prod_i \left({\partial \over \partial\lambda_i} \right)^{n_i} \right)
\ln\langle\phi| {\rm e}^{\xi S^+} \prod_i {\rm e}^{\lambda_i A_i} {\rm e}^{\eta S}|\phi\rangle\,
\right ]_{\xi=\eta=0 \atop \lambda_i=0\,\forall\,i}\,.
\end{eqnarray}
The last expression can be interpreted as a series expansion of the term
in brackets $[...]$ with respect to $\xi$ and $\eta$ around 0:
\begin{eqnarray}
&&\langle\phi| {\rm e}^{\alpha S^+} \prod_i A_i^{n_i} \, {\rm e}^{\beta S} |\phi\rangle^c
\nonumber\\
&&=\:
\left(\prod_i \left({\partial \over \partial\lambda_i} \right)^{n_i} \right)
\ln\langle\phi| {\rm e}^{\alpha S^+} \prod_i {\rm e}^{\lambda_i A_i} {\rm e}^{\beta S}|\phi\rangle\,
|_{\lambda_i=0\,\forall\,i} \nonumber \\
&&=\:
 \langle {\rm e}^{\alpha S} \phi| \,\, \prod_i A_i^{n_i} \,\,
 | {\rm e}^{\beta S} \phi\rangle^c
\,.
\end{eqnarray}
In the last equation we have reintroduced (generalized) cumulants, now formed with
the bra state $\langle {\rm e}^{\alpha S} \phi|$ and the ket state 
$| {\rm e}^{\beta S} \phi\rangle$. With $\alpha=\beta=1$ we obtain
the desired result (\ref{KUMREL1}).

Explicitly we find from (\ref{KUMREL1}):
\begin{eqnarray}
\langle\phi|\,{\rm e}^{S^+} A\, {\rm e}^S\,|\phi\rangle^c\, &=& \,
  { \langle\phi|\,{\rm e}^{S^+} A\, {\rm e}^S\,|\phi\rangle \over
    \langle\phi|\,{\rm e}^{S^+} {\rm e}^S\,|\phi\rangle } \,, \\
\langle\phi|\,{\rm e}^{S^+} AB\, {\rm e}^S\,|\phi\rangle^c\, &=& \,
  { \langle\phi|\,{\rm e}^{S^+} AB\, {\rm e}^S\,|\phi\rangle \over
    \langle\phi|\,{\rm e}^{S^+} {\rm e}^S\,|\phi\rangle }
\nonumber\\
&-&\,
  { \langle\phi|\,{\rm e}^{S^+} A\, {\rm e}^S\,|\phi\rangle \,
    \langle\phi|\,{\rm e}^{S^+} B\, {\rm e}^S\,|\phi\rangle \over
    \langle\phi|\,{\rm e}^{S^+} {\rm e}^S\,|\phi\rangle^2 } \, ,
\nonumber
\end{eqnarray}
and so on.

\end{multicols}


\begin{references}

\bibitem{ExpOptCon}J. Orenstein et $al.$, Phys. Rev. B {\bf 42}, 6342 (1990), \\
	S. Uchida et $al.$, Phys. Rev. B {\bf 43}, 7942 (1991).
\bibitem{Dagotto94}E. Dagotto, Rev. Mod. Phys. {\bf 66}, 763 (1994).
\bibitem{ToHoMae95}T. Tohyama, P. Horsch, and S. Maekawa,  Phys. Rev. Lett. {\bf 74}, 
        980 (1995).
\bibitem{EdOh95}R. Eder and Y. Ohta,  Phys. Rev. B {\bf 51}, 11683 (1995).
\bibitem{EdOhMae95}R. Eder, Y. Ohta, and S. Maekawa,  Phys. Rev. Lett. {\bf 74}, 
        5124 (1995).
\bibitem{EdWroOh96}R. Eder, P. Wrobel, and Y. Ohta,  Phys. Rev. B {\bf 54}, R11034 (1996).
\bibitem{JaPre95}J. Jaklic and P. Prelovsek, Phys. Rev. B {\bf 52}, 6903 (1995).
\bibitem{Kotliar}Z. Wang, Y. Bang, and G. Kotliar,  Phys. Rev. Lett. 
        {\bf 67}, 2733 (1991), \\
        Y. Bang and G. Kotliar, Phys. Rev. B {\bf 48}, 9898 (1993).
\bibitem{GeZey95}L. Gehlhoff and R. Zeyher, Phys. Rev. B {\bf 52}, 4635 (1995).
\bibitem{Lee96}D. K. K. Lee, D. H. Kim, and P. A. Lee, Phys. Rev. Lett. 
        {\bf 76}, 4801 (1996).
\bibitem{KhaHo96}G. Khaliullin and P. Horsch, Phys. Rev. B {\bf 54}, R9600 (1996).
\bibitem{ZeyKul96}R. Zeyher and M. L. Kulic, Phys. Rev. B {\bf 54}, 8985 (1996).
\bibitem{Anderson87}P. W. Anderson, Science {\bf 235}, 1196 (1987).
\bibitem{Zhang88}F. C. Zhang and T. M. Rice, Phys.\ Rev.\ B
        {\bf 37}, 3759 (1988).
\bibitem{Mori}H. Mori, Progr. Theor. Phys. {\bf 34}, 423 (1965).
\bibitem{Zwanzig}R. Zwanzig, in: Lectures in
        Theoretical Physics \ vol. 3. New York: Interscience 1961.
\bibitem{Forster}D. Forster: Hydrodynamic Fluctuations,
        Broken Symmetry, and Correlation Functions. Reading, MA:
        Benjamin 1975.
\bibitem{VojBeck3}M. Vojta and K. W. Becker,  Europhys. Lett. {\bf 38}, 607 (1997).
\bibitem{BeckFul88}K. W. Becker and P. Fulde, Z. Phys. B {\bf 72}, 423 (1988).
\bibitem{BeckWonFul89}K. W. Becker, H. Won, and P. Fulde,
        Z. Phys. B {\bf 75}, 335, (1989).
\bibitem{BeckBre90}K. W. Becker and W. Brenig, Z. Phys. B {\bf 79}, 195 (1990).
\bibitem{Kubo}R. Kubo, J. Phys. Soc. Jpn. {\bf 17}, 1700 (1962).
\bibitem{VojBeck96}M. Vojta and K. W. Becker, Ann. Phys. (Leipzig) {\bf 5},
        156 (1996), \\
        M. Vojta and K. W. Becker, Phys. Rev. B {\bf 54}, 15483 (1996).
\bibitem{SchorkFul92}T. Schork and P. Fulde, J. Chem. Phys. {\bf 97}, 9195 (1992).
\bibitem{Nagaoka}Y. Nagaoka, Phys. Rev. {\bf 147}, 392 (1966).
\bibitem{BrinkRice}W. F. Brinkman and T. M. Rice, Phys. Rev. B {\bf 2}, 1324 (1970).
\bibitem{Trugman88}S. A. Trugman, Phys. Rev. B {\bf 37}, 1597 (1988).
\bibitem{ShrSig88}B. I. Shraiman and E. D. Siggia, Phys. Rev. Lett. {\bf 60}, 740 (1988).
\bibitem{Schrieffer}J. R. Schrieffer, X. G. Wen, and S. C. Zhang, 
                Phys. Rev. Lett {\bf 60}, 944 (1988).
\bibitem{BEW}K. W. Becker, R. Eder, and H. Won, Phys. Rev. B {\bf 45}, 4864 (1992).
\bibitem{Kane}C. L. Kane, P. A. Lee, and N. Read, Phys.  Rev. 
                B {\bf 39}, 6880 (1989).
\bibitem{Kladko}K. Kladko and P. Fulde, Int. J. Quant. Chem. {\bf 66}, 377 (1998).

\end{references}
\end{document}